\documentclass{article}

\usepackage{arxiv}

\usepackage[utf8]{inputenc} 
\usepackage[T1]{fontenc}    
\usepackage{hyperref}       
\usepackage{url}            
\usepackage{booktabs}       
\usepackage{amsfonts}       
\usepackage{amssymb}        
\usepackage{amsmath}        
\usepackage{nicefrac}       
\usepackage{microtype}      
\usepackage{lipsum}		
\usepackage{graphicx}
\usepackage[numbers]{natbib}
\usepackage{doi}
\usepackage{wrapfig}

\title{Fast multivariate polynomials in R: the mvp package}

\author{ \href{https://orcid.org/0000-0001-5982-0415}{\includegraphics[width=0.03\textwidth]{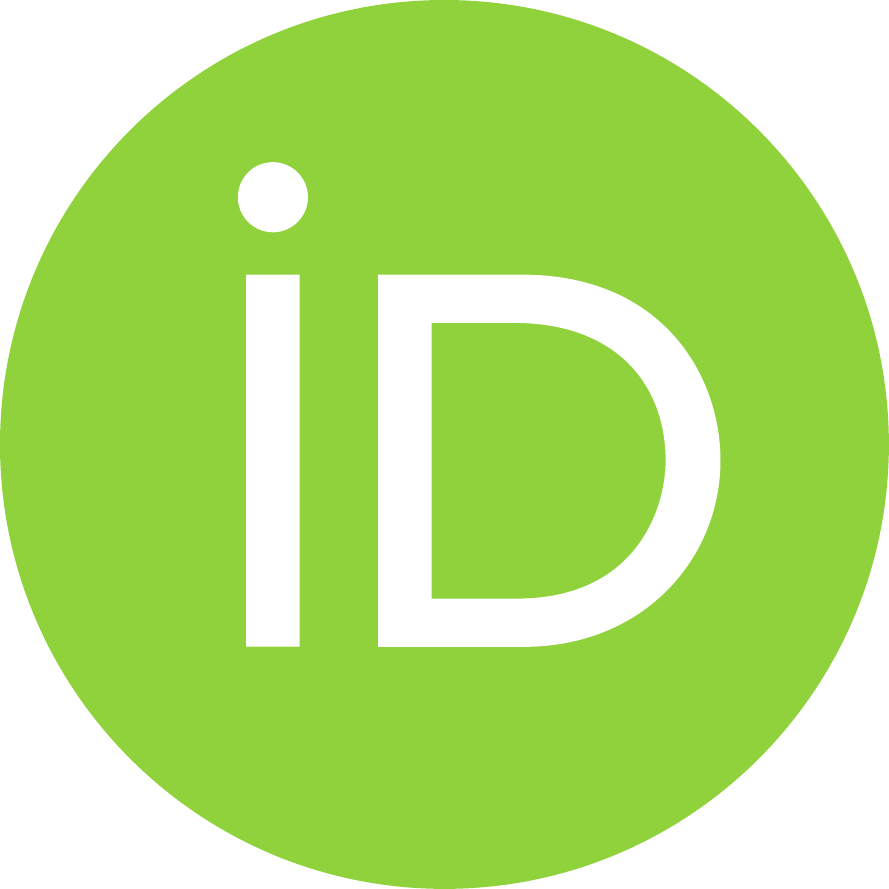}\hspace{1mm}Robin K. S.~Hankin}\thanks{\href{https://academics.aut.ac.nz/robin.hankin}{work};  
\href{https://www.youtube.com/watch?v=JzCX3FqDIOc&list=PL9_n3Tqzq9iWtgD8POJFdnVUCZ_zw6OiB&ab_channel=TrinTragulaGeneralRelativity}{play}} \\
 Auckland University of Technology\\
	\texttt{hankin.robin@gmail.com} \\
}

\renewcommand{\shorttitle}{The \textit{mvp} package}

\hypersetup{
pdftitle={Fast multivariate polynomials in R},
pdfsubject={q-bio.NC, q-bio.QM},
pdfauthor={Robin K. S.~Hankin},
pdfkeywords={multivariate polynomials}
}

\usepackage{Sweave}
\begin{document}
\maketitle

\begin{abstract}
  In this short article I introduce the {\tt mvp} package, which
  provides some functionality for handling multivariate polynomials.
  The package uses the~{\tt C++} Standard Template Library's {\tt map}
  class to store and retrieve elements; it conforms to {\tt disordR}
  discipline for coefficients.  The package is available on CRAN at
  \url{https://CRAN.R-project.org/package=mvp}.
\end{abstract}
\keywords{Multivariate polynomials}

\section{Introduction}

\setlength{\intextsep}{0pt}
\begin{wrapfigure}{r}{0.2\textwidth}
  \begin{center}
\includegraphics[width=1in]{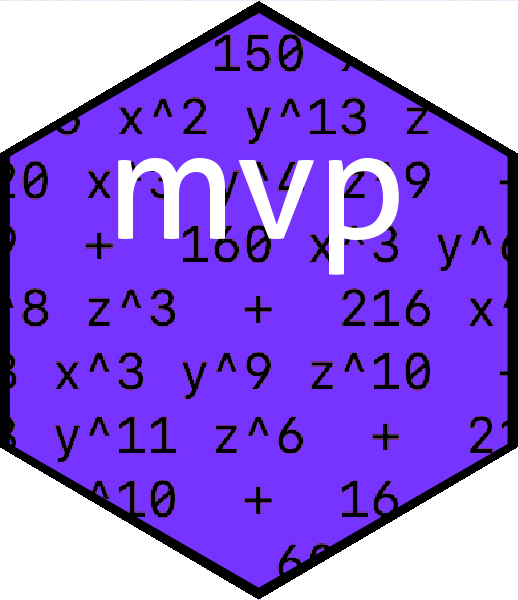}
  \end{center}
\end{wrapfigure}

The {\tt mvp} package provides some functionality for fast
manipulation of multivariate polynomials in the R programming
language~\citep{rcore2022}.  It uses the Standard Template library of
{\tt C++}~\cite{musser2009}, commonly known as the {\tt STL}, and is
comparable in speed to the {\tt spray} package~\cite{hankin2022_spray}
for sparse arrays, while while retaining the symbolic capabilities of
the {\tt mpoly} package~\cite{kahle2013}; the package includes some
timing results.  The {\tt mvp} package uses the excellent print and
coercion methods of {\tt mpoly}.  The {\tt mvp} package provides
improved speed over {\tt mpoly} but retains some of the sophisticated
substitution mechanism.

\section{The {\tt STL} map class}

A ``map'' is a sorted associative container that stores key-value
pairs with unique keys~\citep{musser2009}.  It is interesting here
because search and insertion operations have logarithmic complexity.
Multivariate polynomials are considered to be the sum of a finite
number of {\bf terms}, each multiplied by a coefficient.  A term is
something like $x^2y^3z$.  We may consider this term to be the map

\begin{verbatim}
{"x" -> 2, "y" -> 3, "z" -> 1}
\end{verbatim}

where the map takes symbols to their (integer) power; it is understood
that powers are nonzero.  An {\tt mvp} object is a map from terms to
their coefficients; thus $7xy^2 -3x^2yz^5$ would be

\begin{verbatim}
{{"x" -> 1, "y" -> 2} -> 7, {"x" -> 2, "y" -> 1, "z" ->5} -> -3}
\end{verbatim}

and we understand that coefficients are nonzero.  In {\tt C++} the
declarations would be

\begin{verbatim}
typedef vector <signed int> mypowers;  
typedef vector <string> mynames;  

typedef map <string, signed int> term; 
typedef map <term, double> mvp; 
\end{verbatim}

Thus a term maps a string to a (signed) integer, and a {\tt mvp} maps
terms to doubles.  One reason why the {\tt map} class is fast is that
the order in which the keys are stored is undefined: the compiler may
store them in the order which it regards as most propitious.  This is
not an issue for the maps considered here as addition and
multiplication are commutative and associative.  The package uses {\tt
  disordR} discipline, discussed below in section~\ref{disordsection}.

Note also that constant terms are handled with no difficulty
(constants are simply maps from the empty map to its value), as is the
zero polynomial (which is simply an empty map).

\section{The package in use}

Consider a simple multivariate polynomial $3xy+z^3+xy^6z$ and its
representation in the following R session:

\begin{Schunk}
\begin{Sinput}
R> library("mvp",quietly=TRUE)
R> (p <- as.mvp("3 x y + z^3 + x y^6 z"))
\end{Sinput}
\begin{Soutput}
mvp object algebraically equal to
3 x y + x y^6 z + z^3
\end{Soutput}
\end{Schunk}

Coercion and printing are accomplished by the {\tt mpoly} package
(there is no way I could improve upon
Kahle's work).  Note carefully that the printed representation of
the mvp object is created by the {\tt mpoly} package and the print method
can rearrange both the terms of the polynomial
($3xy+z^3+xy^6z = z^3+3xy+xy^6z$, for example) and the symbols within
a term ($3xy=3yx$, for example) to display the polynomial in a
human-friendly form.

However, note carefully that such rearranging does not affect the
mathematical properties of the polynomial itself.  In the {\tt mvp}
package, the order of the terms is not preserved (or even defined) in
the internal representation of the object; and neither is the order of
the symbols within a single term.  Although this might sound odd, if
we consider a marginally more involved situation, such as

\begin{Schunk}
\begin{Sinput}
R> (M <- as.mvp("3 stoat goat^6 -4 + 7 stoatboat^3 bloat -9 float boat goat gloat^6"))
\end{Sinput}
\begin{Soutput}
mvp object algebraically equal to
-4 + 7 bloat stoatboat^3 - 9 boat float gloat^6 goat + 3 goat^6 stoat
\end{Soutput}
\begin{Sinput}
R> dput(M)
\end{Sinput}
\begin{Soutput}
structure(list(names = list(character(0), c("bloat", "stoatboat"
), c("boat", "float", "gloat", "goat"), c("goat", "stoat")), 
    power = list(integer(0), c(1L, 3L), c(1L, 1L, 6L, 1L), c(6L, 
    1L)), coeffs = c(-4, 7, -9, 3)), class = "mvp")
\end{Soutput}
\end{Schunk}

it is not clear that any human-discernible ordering is preferable to
any other, and we would be better off letting the compiler decide a
propitious ordering.  In any event, the {\tt mpoly} package can
specify a print order:

\begin{Schunk}
\begin{Sinput}
R> print(M,order="lex", varorder=c("stoat","goat","boat","bloat","gloat","float","stoatboat"))
\end{Sinput}
\begin{Soutput}
mvp object algebraically equal to
-4 + 7 bloat stoatboat^3 - 9 boat float gloat^6 goat + 3 goat^6 stoat
\end{Soutput}
\end{Schunk}

Package idiom ({\tt disordR} discipline) for dealing with the
undefined order of terms is discussed below in
section~\ref{disordsection}.

\subsection{Arithmetic operations}

The arithmetic operations {\tt *}, {\tt +}, {\tt -} and \verb|^| work
as expected:

\begin{Schunk}
\begin{Sinput}
R> (S1 <- rmvp(5,2,2,4))
\end{Sinput}
\begin{Soutput}
mvp object algebraically equal to
b d^2 + 5 b^2 d^2 + 2 b^4 + 4 c d + 3 c d^2
\end{Soutput}
\begin{Sinput}
R> (S2 <- rmvp(5,2,2,4))
\end{Sinput}
\begin{Soutput}
mvp object algebraically equal to
2 b^4 + 4 c^2 d + 5 c^4 + 4 d^4
\end{Soutput}
\begin{Sinput}
R> S1 + S2
\end{Sinput}
\begin{Soutput}
mvp object algebraically equal to
b d^2 + 5 b^2 d^2 + 4 b^4 + 4 c d + 3 c d^2 + 4 c^2 d + 5 c^4 + 4 d^4
\end{Soutput}
\begin{Sinput}
R> S1 * S2
\end{Sinput}
\begin{Soutput}
mvp object algebraically equal to
4 b c^2 d^3 + 5 b c^4 d^2 + 4 b d^6 + 20 b^2 c^2 d^3 + 25 b^2 c^4 d^2
+ 20 b^2 d^6 + 8 b^4 c d + 6 b^4 c d^2 + 8 b^4 c^2 d + 10 b^4 c^4 + 8
b^4 d^4 + 2 b^5 d^2 + 10 b^6 d^2 + 4 b^8 + 16 c d^5 + 12 c d^6 + 16
c^3 d^2 + 12 c^3 d^3 + 20 c^5 d + 15 c^5 d^2
\end{Soutput}
\begin{Sinput}
R> S1^2
\end{Sinput}
\begin{Soutput}
mvp object algebraically equal to
8 b c d^3 + 6 b c d^4 + 40 b^2 c d^3 + 30 b^2 c d^4 + b^2 d^4 + 10
b^3 d^4 + 16 b^4 c d + 12 b^4 c d^2 + 25 b^4 d^4 + 4 b^5 d^2 + 20 b^6
d^2 + 4 b^8 + 16 c^2 d^2 + 24 c^2 d^3 + 9 c^2 d^4
\end{Soutput}
\end{Schunk}

\subsection{Substitution}

The package has two substitution functionalities.  Firstly, we can
substitute one or more variables for a numeric value.  Define a mvp
object:

\begin{Schunk}
\begin{Sinput}
R> (S3 <- as.mvp("x + 5 x^4 y + 8 y^2 x z^3"))
\end{Sinput}
\begin{Soutput}
mvp object algebraically equal to
x + 8 x y^2 z^3 + 5 x^4 y
\end{Soutput}
\end{Schunk}

And then we may substitute $x=1$:

\begin{Schunk}
\begin{Sinput}
R> subs(S3, x = 1)
\end{Sinput}
\begin{Soutput}
mvp object algebraically equal to
1 + 5 y + 8 y^2 z^3
\end{Soutput}
\end{Schunk}

Note the natural R idiom, and that the return value is another mvp
object.  We may substitute for the other variables:

\begin{Schunk}
\begin{Sinput}
R> subs(S3, x = 1, y = 2, z = 3)
\end{Sinput}
\begin{Soutput}
[1] 875
\end{Soutput}
\end{Schunk}

(in this case, the default behaviour is to return the the resulting
polynomial coerced to a scalar).  We can suppress the coercion using
the {\tt lose} argument:

\begin{Schunk}
\begin{Sinput}
R> subs(S3, x = 1, y = 2, z = 3,lose=FALSE)
\end{Sinput}
\begin{Soutput}
mvp object algebraically equal to
875
\end{Soutput}
\end{Schunk}

The idiom also allows one to substitute a variable for an {\tt mvp}
object:

\begin{Schunk}
\begin{Sinput}
R> subs(as.mvp("a+b+c"), a="x^6")
\end{Sinput}
\begin{Soutput}
mvp object algebraically equal to
b + c + x^6
\end{Soutput}
\end{Schunk}

Note carefully that {\tt subs()} depends on the order of substitution:

\begin{Schunk}
\begin{Sinput}
R> subs(as.mvp("a+b+c"), a="x^6",x="1+a")
\end{Sinput}
\begin{Soutput}
mvp object algebraically equal to
1 + 6 a + 15 a^2 + 20 a^3 + 15 a^4 + 6 a^5 + a^6 + b + c
\end{Soutput}
\begin{Sinput}
R> subs(as.mvp("a+b+c"), x="1+a",a="x^6")
\end{Sinput}
\begin{Soutput}
mvp object algebraically equal to
b + c + x^6
\end{Soutput}
\end{Schunk}

\subsection{Pipes}

Substitution works well with pipes:

\begin{Schunk}
\begin{Sinput}
R> as.mvp("a+b") %>% subs(a="a^2+b^2") %>% subs(b="x^6")
\end{Sinput}
\begin{Soutput}
mvp object algebraically equal to
a^2 + x^6 + x^12
\end{Soutput}
\end{Schunk}

\subsection{Vectorised substitution}

Function {\tt subvec()} allows one to substitute variables for numeric
values using vectorised idiom:

\begin{Schunk}
\begin{Sinput}
R> p <- rmvp(6,2,2,letters[1:3])
R> p
\end{Sinput}
\begin{Soutput}
mvp object algebraically equal to
3 a c + 6 a^2 b^2 + 8 a^2 c^2 + 4 a^4
\end{Soutput}
\begin{Sinput}
R> subvec(p,a=1,b=2,c=1:5)   # supply a named list of vectors
\end{Sinput}
\begin{Soutput}
[1]  39  66 109 168 243
\end{Soutput}
\end{Schunk}

\subsection{Differentiation}

Differentiation is implemented.  First we have the {\tt deriv()}
method:

\begin{Schunk}
\begin{Sinput}
R> (S <- as.mvp("a + 5 a^5*b^2*c^8 -3 x^2 a^3 b c^3"))
\end{Sinput}
\begin{Soutput}
mvp object algebraically equal to
a - 3 a^3 b c^3 x^2 + 5 a^5 b^2 c^8
\end{Soutput}
\begin{Sinput}
R> deriv(S, letters[1:3])
\end{Sinput}
\begin{Soutput}
mvp object algebraically equal to
-27 a^2 c^2 x^2 + 400 a^4 b c^7
\end{Soutput}
\begin{Sinput}
R> deriv(S, rev(letters[1:3]))  # should be the same.
\end{Sinput}
\begin{Soutput}
mvp object algebraically equal to
-27 a^2 c^2 x^2 + 400 a^4 b c^7
\end{Soutput}
\end{Schunk}

Also a slightly different form: {\tt aderiv()}, here used to evaluate
$\frac{\partial^6S}{\partial a^3\partial b\partial c^2}$:

\begin{Schunk}
\begin{Sinput}
R> aderiv(S, a = 3, b = 1, c = 2)
\end{Sinput}
\begin{Soutput}
mvp object algebraically equal to
33600 a^2 b c^6 - 108 c x^2
\end{Soutput}
\end{Schunk}

Again, pipes work quite nicely:

\begin{Schunk}
\begin{Sinput}
R> S %<>% aderiv(a=1,b=2) %>% subs(c="x^4") %>% `+`(as.mvp("o^99"))
R> S
\end{Sinput}
\begin{Soutput}
mvp object algebraically equal to
50 a^4 x^32 + o^99
\end{Soutput}
\end{Schunk}

\subsection{Taylor series}

The package includes functionality to deal with Taylor and Laurent
series.  The {\tt trunc()} function allows one to truncate a
polynomial, retain only terms with total power less than a particular
integer, for example:

\begin{Schunk}
\begin{Sinput}
R> (X <- as.mvp("1+x+x^2 y")^3)
\end{Sinput}
\begin{Soutput}
mvp object algebraically equal to
1 + 3 x + 3 x^2 + 3 x^2 y + x^3 + 6 x^3 y + 3 x^4 y + 3 x^4 y^2 + 3
x^5 y^2 + x^6 y^3
\end{Soutput}
\begin{Sinput}
R> trunc(X,3)        
\end{Sinput}
\begin{Soutput}
mvp object algebraically equal to
1 + 3 x + 3 x^2 + 3 x^2 y + x^3
\end{Soutput}
\end{Schunk}

Alternatively, we could retain only terms with powers of $x$ {\em
  less} than a particular value:

\begin{Schunk}
\begin{Sinput}
R> trunc1(X,x=3)    # retain only terms with power of x <= 3
\end{Sinput}
\begin{Soutput}
mvp object algebraically equal to
1 + 3 x + 3 x^2 + 3 x^2 y + x^3 + 6 x^3 y
\end{Soutput}
\end{Schunk}

If we wish to return powers of $x$ that are {\em equal} to a
particular value, we need to use {\tt onevarpow()}:

\begin{Schunk}
\begin{Sinput}
R> onevarpow(X,x=3) # retain only terms with power of x == 3
\end{Sinput}
\begin{Soutput}
mvp object algebraically equal to
1 + 6 y
\end{Soutput}
\end{Schunk}

The ideas above allow us to express Taylor series in package idiom.
For example, consider $\sin_5(x+y)$, where $\sin_n$ is shorthand for
the fifth-order Taylor expansion of $\sin$.  We wish to evaluate the
second order Taylor expansion of $f(x,y)=\sin_5(x+y)$ about $x=1.1$.
We define a {\tt mvp} object which is correct to fifth order:

\begin{Schunk}
\begin{Sinput}
R> (sinxpy <- horner("x+y",c(0,1,0,-1/6,0,+1/120)))  # sin(x+y)
\end{Sinput}
\begin{Soutput}
mvp object algebraically equal to
x - 0.5 x y^2 + 0.04166667 x y^4 - 0.5 x^2 y + 0.08333333 x^2 y^3 -
0.1666667 x^3 + 0.08333333 x^3 y^2 + 0.04166667 x^4 y + 0.008333333
x^5 + y - 0.1666667 y^3 + 0.008333333 y^5
\end{Soutput}
\end{Schunk}

and then define an object {\tt dx} with the intent that this is
``small"

\begin{Schunk}
\begin{Sinput}
R> dx <- as.mvp("dx")
\end{Sinput}
\end{Schunk}

which would allow us to create a {\em symbolic} Taylor series to
second order using {\tt aderiv()}:

\begin{Schunk}
\begin{Sinput}
R> (t2 <- sinxpy  + aderiv(sinxpy,x=1)*dx + aderiv(sinxpy,x=2)*dx^2/2)
\end{Sinput}
\begin{Soutput}
mvp object algebraically equal to
dx - dx x y + 0.1666667 dx x y^3 - 0.5 dx x^2 + 0.25 dx x^2 y^2 +
0.1666667 dx x^3 y + 0.04166667 dx x^4 - 0.5 dx y^2 + 0.04166667 dx
y^4 - 0.5 dx^2 x + 0.25 dx^2 x y^2 + 0.25 dx^2 x^2 y + 0.08333333
dx^2 x^3 - 0.5 dx^2 y + 0.08333333 dx^2 y^3 + x - 0.5 x y^2 +
0.04166667 x y^4 - 0.5 x^2 y + 0.08333333 x^2 y^3 - 0.1666667 x^3 +
0.08333333 x^3 y^2 + 0.04166667 x^4 y + 0.008333333 x^5 + y -
0.1666667 y^3 + 0.008333333 y^5
\end{Soutput}
\end{Schunk}

Noting that the terms are stored in an implementation-specific order,
above we see an {\tt mvp} object equal to the right hand side of

\[
\sin(x+y+\delta x)\simeq
\sin(x+y) + 
 \delta x\cdot\frac{\partial}{\partial x}\sin(x+y) +
\frac{(\delta x)^2}{2!}\cdot\frac{\partial^2}{\partial x^2}\sin(x+y) 
\]

(the second argument of function {\tt aderiv()} gives the order of
differentiation).  To evaluate the second order Taylor expansion of
$\sin(y+1.1)$, left in symbolic form we would substitute for the
numerical value of {\tt x} and {\tt dx} which we will take to be 0.1:

\begin{Schunk}
\begin{Sinput}
R> (t2 %<>% subs(x=1.1, dx=0.1))
\end{Sinput}
\begin{Soutput}
mvp object algebraically equal to
0.9327972 + 0.3662125 y - 0.4560833 y^2 - 0.04666667 y^3 + 0.05 y^4 +
0.008333333 y^5
\end{Soutput}
\end{Schunk}

See that this Taylor expansion retains a symbolic dependence on $y$.
For numerical verification we will evaluate this at $y=0.3$:

\begin{Schunk}
\begin{Sinput}
R> (t2 %>% subs(y=0.3))
\end{Sinput}
\begin{Soutput}
[1] 1.000779
\end{Soutput}
\begin{Sinput}
R> t2 %>% subs(y=0.3)  - sinxpy %>% subs(x=1.1,y=0.4)
\end{Sinput}
\begin{Soutput}
[1] -2.583333e-06
\end{Soutput}
\end{Schunk}

Above we see that the error in evaluating
$\sin_5(1.5)=\sin_5(1.1+0.3+\delta x)$ using a second order Taylor
expansion in $\delta x=0.1$ about $\sin_5(1.4)$ is ${\mathcal
O}(\delta x)^3\sim 10^{-3}$.  The package is thus seen to be capable
of representing objects with differing degrees of symbolic and numeric
constitution.

\subsection{Power series}

Function {\tt series()} will decompose an {\tt mvp} object into a power
series in a single variable:

\begin{Schunk}
\begin{Sinput}
R> p <- as.mvp("a^2 x b + x^2 a b + b c x^2 + a b c + c^6 x")
R> p
\end{Sinput}
\begin{Soutput}
mvp object algebraically equal to
a b c + a b x^2 + a^2 b x + b c x^2 + c^6 x
\end{Soutput}
\begin{Sinput}
R> series(p,"x")
\end{Sinput}
\begin{Soutput}
x^0(a b c)  + x^1(a^2 b  +  c^6)  + x^2(a b  +  b c)
\end{Soutput}
\end{Schunk}

This works nicely with {\tt subs()} if we wish to take a power series
about {\tt x-v}, where {\tt v} is any {\tt mvp} object.  For example:

\begin{Schunk}
\begin{Sinput}
R> p %>% subs(x="xmv+a+b") %>% series("xmv") 
\end{Sinput}
\begin{Soutput}
xmv^0(a b c  +  2 a b^2 c  +  a b^3  +  a c^6  +  a^2 b c  +  3 a^2 
b^2  +  2 a^3 b  +  b c^6  +  b^3 c)  + xmv^1(2 a b c  +  2 a b^2  +  
3 a^2 b  +  2 b^2 c  +  c^6)  + xmv^2(a b  +  b c)
\end{Soutput}
\end{Schunk}

is a series in powers of {\tt x-a-b}.  We may perform a consistency
check by a second substitution, returning us to the original
expression:

\begin{Schunk}
\begin{Sinput}
R> p == p %>% subs(x="xmv+a+b") %>% subs(xmv="x-a-b")
\end{Sinput}
\begin{Soutput}
[1] TRUE
\end{Soutput}
\end{Schunk}

If function {\tt series()} is given a variable name ending in {\tt
  \_m\_foo}, where {\tt foo} is any variable name, then this is
typeset as {\tt (x-foo)}.  For example:

\begin{Schunk}
\begin{Sinput}
R> as.mvp("x^3 + x*a") %>% subs(x="x_m_a + a") %>% series("x_m_a")
\end{Sinput}
\begin{Soutput}
(x-a)^0(a^2  +  a^3)  + (x-a)^1(a  +  3 a^2)  + (x-a)^2(3 a)  + 
(x-a)^3(1)
\end{Soutput}
\end{Schunk}

So above we see the expansion of $x^2+ax$ in powers of $x-a$.  If we
want to see the expansion of an {\tt mvp} object in terms of a more
complicated expression then it is better to use a nonce variable {\tt
  v}:

\begin{Schunk}
\begin{Sinput}
R> as.mvp("x^2 + x*a+b^3") %>% subs(x="x_m_v + a^2+b") %>% series("x_m_v")
\end{Sinput}
\begin{Soutput}
(x-v)^0(a b  +  2 a^2 b  +  a^3  +  a^4  +  b^2  +  b^3)  + (x-v)^1(a 
 +  2 a^2  +  2 b)  + (x-v)^2(1)
\end{Soutput}
\end{Schunk}

where it is understood that $v=a+b^2$.  Function {\tt taylor()} is a
convenience wrapper that does some of the above in one step:

\begin{Schunk}
\begin{Sinput}
R> p <- as.mvp("1+x-x*y+a")^2
R> taylor(p,"x","a")
\end{Sinput}
\begin{Soutput}
(x-a)^0(1  +  4 a  -  2 a y  +  4 a^2  -  4 a^2 y  +  a^2 y^2)  + 
(x-a)^1(2  +  4 a  -  6 a y  +  2 a y^2  -  2 y)  + (x-a)^2(1  -  2 y 
 +  y^2)
\end{Soutput}
\end{Schunk}

However, this functionality is under development.

\subsection{Extraction}

Given a multivariate polynomial, one often needs to extract certain
terms.  Because the terms of an {\tt mvp} object have an
implementation-dependent order, this can be difficult.  But we can use
function {\tt onevarpow()}:

\begin{Schunk}
\begin{Sinput}
R> (P <- as.mvp("1 + z + y^2 + x*z^2 +  x*y")^3)
\end{Sinput}
\begin{Soutput}
mvp object algebraically equal to
1 + 3 x y + 6 x y z + 3 x y z^2 + 6 x y^2 z^2 + 6 x y^2 z^3 + 6 x y^3
+ 6 x y^3 z + 3 x y^4 z^2 + 3 x y^5 + 3 x z^2 + 6 x z^3 + 3 x z^4 + 6
x^2 y z^2 + 6 x^2 y z^3 + 3 x^2 y^2 + 3 x^2 y^2 z + 3 x^2 y^2 z^4 + 6
x^2 y^3 z^2 + 3 x^2 y^4 + 3 x^2 z^4 + 3 x^2 z^5 + 3 x^3 y z^4 + 3 x^3
y^2 z^2 + x^3 y^3 + x^3 z^6 + 3 y^2 + 6 y^2 z + 3 y^2 z^2 + 3 y^4 + 3
y^4 z + y^6 + 3 z + 3 z^2 + z^3
\end{Soutput}
\begin{Sinput}
R> onevarpow(P,x=1,y=2)
\end{Sinput}
\begin{Soutput}
mvp object algebraically equal to
6 z^2 + 6 z^3
\end{Soutput}
\end{Schunk}

Above we see function {\tt onevarpow()} returns an {\tt mvp} object
including only those terms with $x$ to the first power and $y$ to the
second power (the factor of $x^1y^2$ is dropped from the result).

\subsection{Negative powers}

The {\tt mvp} package handles negative powers, although the idiom is
not perfect and is still under development.  Consider experimental
function {\tt invert()}:

\begin{Schunk}
\begin{Sinput}
R> (p <- as.mvp("1+x+x^2 y"))
\end{Sinput}
\begin{Soutput}
mvp object algebraically equal to
1 + x + x^2 y
\end{Soutput}
\begin{Sinput}
R> invert(p)
\end{Sinput}
\begin{Soutput}
mvp object algebraically equal to
1 + x^-2 y^-1 + x^-1
\end{Soutput}
\end{Schunk}

In the above, {\tt p} is a regular multivariate polynomial which
includes negative powers.  It obeys the same arithmetic rules as other
mvp objects:

\begin{Schunk}
\begin{Sinput}
R> p + as.mvp("z^6")
\end{Sinput}
\begin{Soutput}
mvp object algebraically equal to
1 + x + x^2 y + z^6
\end{Soutput}
\end{Schunk}

\section{The {\tt disordR} \label{disordsection} package}

As discussed above, the terms of an {\tt mvp} object are held in an
implementation-specific order and dealing with this in a consistent
way is achieved using the {\tt disordR}
package~\cite{hankin2022_disordR}.  It is possible to examine the
coefficients of an {\tt mvp} object:

\begin{Schunk}
\begin{Sinput}
R> a <- as.mvp("5 + 8*x^2*y - 13*y*x^2 + 11*z - 3*x*yz")
R> a
\end{Sinput}
\begin{Soutput}
mvp object algebraically equal to
5 - 3 x yz - 5 x^2 y + 11 z
\end{Soutput}
\begin{Sinput}
R> coeffs(a)
\end{Sinput}
\begin{Soutput}
A disord object with hash 64e403dc35ccd68b1bcbd3e0444c72f4e57b50fd and elements
[1]  5 -3 -5 11
(in some order)
\end{Soutput}
\end{Schunk}

Above, note that the result of {\tt coeffs()} is a {\tt disord}
object, defined in the {\tt disordR}
package~\cite{hankin2022_disordR}.  The order of the elements is
unspecified as the {\tt STL} map class holds the keys and values in an
implementation-specific order.  This device stops the user from
illegal operations on the coefficients.  For example, suppose we had
another mvp object, {\tt b}:

\begin{Schunk}
\begin{Sinput}
R> b <- a*2
R> b
\end{Sinput}
\begin{Soutput}
mvp object algebraically equal to
10 - 6 x yz - 10 x^2 y + 22 z
\end{Soutput}
\end{Schunk}

Then we could not add the coefficients of the two objects

\begin{Schunk}
\begin{Sinput}
R> coeffs(a) + coeffs(b)
\end{Sinput}
\begin{Soutput}
Error in check_matching_hash(e1, e2, match.call()) : 
hash codes 64e403dc35ccd68b1bcbd3e0444c72f4e57b50fd and d4875b2ede120f28db25aa644df3ea95f5be7d58 do not match
\end{Soutput}
\end{Schunk}

Above, we get an error because the coefficients of {\tt a} and {\tt b}
are possibly stored in a different order and therefore vector
addition makes no sense.  However, we can operate on coefficients of
a single {\tt mvp} object at will:

\begin{Schunk}
\begin{Sinput}
R> coeffs(a) > 0
\end{Sinput}
\begin{Soutput}
A disord object with hash 64e403dc35ccd68b1bcbd3e0444c72f4e57b50fd and elements
[1]  TRUE FALSE FALSE  TRUE
(in some order)
\end{Soutput}
\begin{Sinput}
R> coeffs(a) + coeffs(a)^4
\end{Sinput}
\begin{Soutput}
A disord object with hash 64e403dc35ccd68b1bcbd3e0444c72f4e57b50fd and elements
[1]   630    78   620 14652
(in some order)
\end{Soutput}
\end{Schunk}

Extraction also works but subject to standard {\tt disordR} idiom
restrictions:

\begin{Schunk}
\begin{Sinput}
R> coeffs(a)[coeffs(a) > 0]
\end{Sinput}
\begin{Soutput}
A disord object with hash d920d86b5eaa22c6a8d3f4ae09bc13feeb87ede4 and elements
[1]  5 11
(in some order)
\end{Soutput}
\end{Schunk}

But ``mixing" objects is forbidden:

\begin{verbatim}
coeffs(a)[coeffs(b) > 0]
\end{verbatim}

Extraction methods work, again subject to {\tt disordR} restrictions:

\begin{Schunk}
\begin{Sinput}
R> coeffs(a)[coeffs(a)<0] <- coeffs(a)[coeffs(a)<0] + 1000 # add 1000 to every negative coefficient
R> a
\end{Sinput}
\begin{Soutput}
mvp object algebraically equal to
5 + 997 x yz + 995 x^2 y + 11 z
\end{Soutput}
\end{Schunk}

In cases like this where the replacement object is complicated, using
{\tt magrittr} would simplify the idiom and reduce the opportunity for
error:

\begin{Schunk}
\begin{Sinput}
R> library("magrittr")
R> b
\end{Sinput}
\begin{Soutput}
mvp object algebraically equal to
10 - 6 x yz - 10 x^2 y + 22 z
\end{Soutput}
\begin{Sinput}
R> coeffs(b)[coeffs(b)%%3==1] %<>% `+`(100)  # add 100 to every element equal to 1 modulo 3
R> b
\end{Sinput}
\begin{Soutput}
mvp object algebraically equal to
110 - 6 x yz - 10 x^2 y + 122 z
\end{Soutput}
\end{Schunk}

One good use for this is to ``zap" small elements:

\begin{Schunk}
\begin{Sinput}
R> x <- as.mvp("1 - 0.11*x + 0.005*x*y")^2
R> x
\end{Sinput}
\begin{Soutput}
mvp object algebraically equal to
1 - 0.22 x + 0.01 x y + 0.0121 x^2 - 0.0011 x^2 y + 0.000025 x^2 y^2
\end{Soutput}
\end{Schunk}

Then we can zap as follows:

\begin{Schunk}
\begin{Sinput}
R> cx <- coeffs(x)
R> cx[abs(cx) < 0.01] <- 0
R> coeffs(x) <- cx
R> x
\end{Sinput}
\begin{Soutput}
mvp object algebraically equal to
1 - 0.22 x + 0.01 x y + 0.0121 x^2
\end{Soutput}
\end{Schunk}

\section{Example: multivariate generating functions}

Here I give an example of the {\tt mvp} package in use,
following~\cite{hankin2022_spray}.  The generating function for a
chess knight is given by function {\tt knight()}, whose argument is
the dimension of the board.  Thus for a 2D board we have

\begin{Schunk}
\begin{Sinput}
R> knight(2)
\end{Sinput}
\begin{Soutput}
mvp object algebraically equal to
a^-2 b^-1 + a^-2 b + a^-1 b^-2 + a^-1 b^2 + a b^-2 + a b^2 + a^2 b^-1
+ a^2 b
\end{Soutput}
\end{Schunk}

\begin{figure}[p]
  \centering \includegraphics[width=10cm]{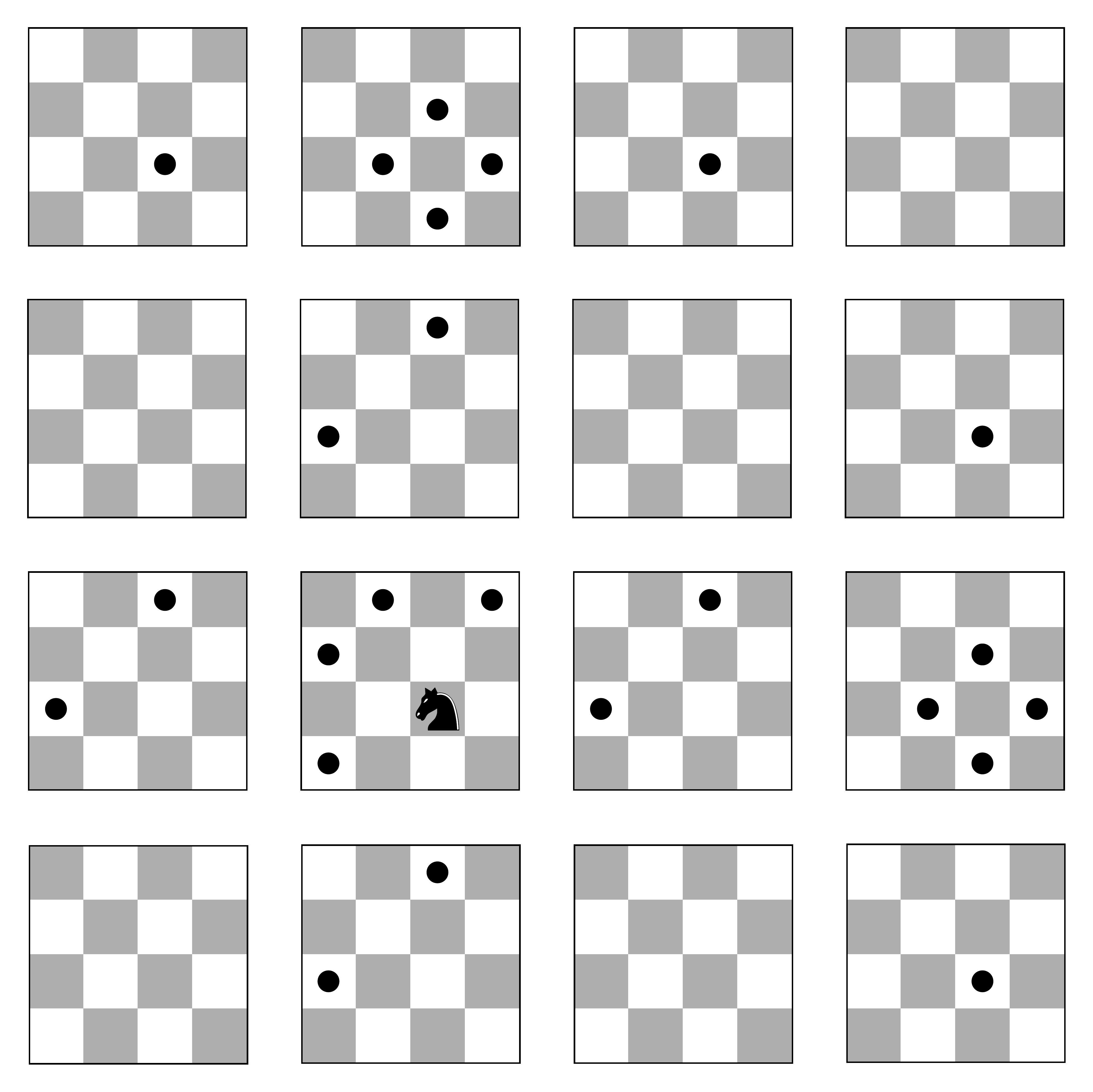}
  \caption{Four-dimensional knight\label{four_dimensional_knight} on a $4\times 4\times 4\times 4$ board.  Cells attacked by the knight shown by dots}
\end{figure}

This object has 8 terms, one for each move that a knight has on an
unrestricted board.  The powers of {\tt a} and {\tt b} correspond to
horizontal and vertical directions respectively.  Now consider a
four-dimensional chessboard (Figure~\ref{four_dimensional_knight}
shows a $4\times 4\times 4\times 4$ board).  How many ways are there
for a 4D knight to return to its starting square after four moves on
an infinite board?  Answer:

\begin{Schunk}
\begin{Sinput}
R> constant(knight(4)^4)
\end{Sinput}
\begin{Soutput}
[1] 12528
\end{Soutput}
\end{Schunk}

We might ask how many ways there are for the knight to move one space
in each of the 4 directions, by extracting the coefficient of $abcd$:

\begin{Schunk}
\begin{Sinput}
R> K <- knight(4)^4
R> onevarpow(K,a=1,b=1,c=1,d=1)
\end{Sinput}
\begin{Soutput}
mvp object algebraically equal to
4536
\end{Soutput}
\end{Schunk}

Further, we might ask which places the knight can move to on the plane
$a=4$, $b=5$:

\begin{Schunk}
\begin{Sinput}
R> onevarpow(K,a=4,b=5)
\end{Sinput}
\begin{Soutput}
mvp object algebraically equal to
60 c^-3 + 96 c^-2 d^-1 + 96 c^-2 d + 192 c^-1 + 96 c^-1 d^-2 + 96
c^-1 d^2 + 192 c + 96 c d^-2 + 96 c d^2 + 96 c^2 d^-1 + 96 c^2 d + 60
c^3 + 60 d^-3 + 192 d^-1 + 192 d + 60 d^3
\end{Soutput}
\end{Schunk}

Thus there are exactly 60 ways for the knight to move to $(4,5,-3,0)$,
96 ways to move to $(4,5,-2,-1)$, and so on.  As a final example, we
might assume that the knight moves randomly, and wish to determine the
probability mass function of its position.  What is its expected
Euclidean distance from the origin after four moves?

\begin{Schunk}
\begin{Sinput}
R> dist <- function(x){sqrt(sum(x^2))}  # regular Euclidean distance function
R> sum(unlist(lapply(powers(K),dist))*coeffs(K))/sum(coeffs(K))
\end{Sinput}
\begin{Soutput}
[1] 4.248183
\end{Soutput}
\end{Schunk}

\bibliographystyle{plain}
\bibliography{mvp_arxiv}
\end{document}